\documentclass{ws-procs975x65}
\begin{document}

\title{Spin, Bose, and Non-Fermi Liquid Metals in Two Dimensions:\\
Accessing via Multi-Leg Ladders}

\author{Matthew P.A. Fisher}

\address{Microsoft Research, Station Q, University of California,\\
Santa Barbara, California 93106, USA \\
E-mail: mpaf@kitp.ucsb.edu\\
}

\author{Olexei I. Motrunich}

\address{Department of Physics, California Institute of Technology\\
Pasadena, CA 91125, USA\\
E-mail: motrunch@caltech.edu}

\author{Donna N. Sheng}

\address{California State University, Northridge\\
Northridge, CA 91330, USA\\
E-mail: donna.sheng@csun.edu}

\begin{abstract}
Characterizing and accessing quantum phases of itinerant 
bosons or fermions in two dimensions (2D) that exhibit 
singular structure along surfaces in momentum space but 
have no quasi-particle description remains as a central 
challenge in the field of strongly correlated physics.
Fortuitously, signatures of such 2D strongly correlated phases 
are expected to be manifest in quasi-one-dimensional 
``$N$-leg ladder" systems.
The ladder discretization of the transverse momentum cuts through 
the 2D surface, leading to a quasi-1D descendant state with a 
set of low-energy modes whose number grows with the number of legs 
and whose momenta are inherited from the 2D surfaces.   
These multi-mode quasi-1D liquids constitute a new and previously 
unanticipated class of quantum states interesting in their own right.  
But more importantly they carry a distinctive quasi-1D ``fingerprint" 
of the parent 2D quantum fluid state.
This can be exploited to access the 2D phases from controlled 
numerical and analytical studies in quasi-1D models. 
The preliminary successes and future prospects in this endeavor 
will be briefly summarized.
\end{abstract}


\bodymatter

\section{Challenges in Mott materials}\label{aba:sec1}

The quantum theory of metals which identifies the Landau quasiparticles 
formed out of Bloch electrons as the appropriate independent 
electron-like excitations is a hallmark of the 20th century physics.
Despite its remarkable successes in describing many (weakly correlated) 
materials, it is now clear that this approach can fail qualitatively 
when the interactions are strong, as often occurs in materials with 
narrow partially filled bands originating from well-localized atomic 
$d$- and $f$-shells.
Most dramatic are materials with a half-filled band that are 
Mott insulators because of the strong local Coulomb repulsion.
An intriguing possibility is that such Mott insulators can exhibit 
exotic spin liquid ground states, having no magnetic 
or any other order\cite{Anderson1,Anderson2}.
A recent breakthrough is the appearance of several experimental 
realizations of spin liquids which all appear to be gapless, 
most notably some transition metal ($d$-shell) Kagome based crystals 
and a class of crystalline organic Mott insulators.  
The heavy fermion materials and cuprate superconductors are 
itinerant electron system which also appear to fall outside the 
rubric of the conventional theory of metals.

Many models have been proposed to understand these systems, such as 
Hubbard, $t-J$, and Kondo lattice models, but we essentially do not 
have controlled approaches to study them.  On the analytical side, 
mean field treatments can look for states with broken symmetries, 
e.g.\ with spin or charge order, but cannot access non-Fermi-liquid 
physics.  Among non-perturbative approaches, one is slave particle 
construction and gauge theory analysis while another is duality 
where one thinks in terms of topological defects like vortices; 
these approaches suggest the possibility of new phases in principle, 
but are uncontrolled for almost all 2D and 3D realistic models. 

On the numerical side, Exact Diagonalization (ED) is limited to
small systems, often too small to extract the physics. 
Quantum Monte Carlo suffers from sign problems. 
Variational Monte Carlo (VMC) calculations suffer from bias in the 
trial states.
Dynamical Mean Field Theory does not capture all the important
spatial correlation physics. 
Density Functional Theory, which is at the heart of realistic 
band structure calculations, describes well the ``high-energy'' 
(core) electrons, but does not capture properly the local Coulomb 
repulsion for the relevant electrons near the Fermi level.
The Density Matrix Renormalization Group (DMRG) works extremely well 
in 1D, but capturing the entanglement inherent in strongly correlated 
2D phases appears daunting.

\section{2D Spin and Bose Metals}

Many exotic spin liquid phases have been suggested by effective 
field theories (mostly gauge theory) and we now know that there are 
different kinds of spin liquids\cite{LeeNagaosaWen}.  
Gapped topological spin liquids are best understood and have been shown 
to exist in model systems\cite{RSSpN, Wen, topth, MoeSon}. 
Gapless spin liquids are also possible and will generically exhibit 
spin correlations that decay as a power law in space, perhaps with 
anomalous exponents, and which can oscillate at particular wavevectors.
The location of these dominant singularities in momentum space provides 
a convenient characterization of gapless spin liquids.
In the  ``algebraic" or ``critical" spin liquids
\cite{WenPSG, Rantner, Hermele, LeeNagaosaWen} these wavevectors are 
limited to a finite discrete set, often at high symmetry points in the 
Brillouin zone.  But the singularities can occur along {\it surfaces} 
in momentum space, as they do in the Gutzwiller projected 
spinon Fermi sea state\cite{Baskaran, LeeNagaosa, LeeNagaosaWen}.    
While the singular surfaces in such ``quantum spin metals" are 
reminiscent of the Fermi surface in a 2D Fermi liquid, it must be 
stressed that it is the {\it spin} correlation functions that possess 
such singular surfaces -- there are no Fermions in the theory -- 
and the low energy excitations cannot be described in terms of 
weakly interacting quasiparticles. 

There has been much less theoretical progress on non-FL conductors. 
Typically, the effective field theories have treated the electron 
charge sector as exhibiting conventional or classical physics. 
To explore the possibility of novel quantum behavior of itinerant charge 
carriers, two of us recently studied a closely related possibility of 
uncondensed but conducting quantum states of bosons\cite{DBL}. 
This work proposed a 2D model of bosons with frustrating ring exchanges 
to realize a novel D-wave Bose Liquid (DBL), a ``Bose metal'' phase 
with low-energy excitations residing on ``Bose surfaces'' in 
momentum space. 
By combining with the spin sector, this can be extended to construct 
non-Fermi-Liquid electron states which have singular surfaces in 
momentum space that violate Luttinger's theorem.  Other examples with 
critical surfaces have been studied recently\cite{Senthil}.

\section{New Quasi-1D approach to Spin and Bose metals}
 
Recently we argued that 2D spin metals, Bose metals, and 
non-Fermi-liquids phases which exhibit many low-energy excitations 
residing on surfaces in momentum space, should be accessible by 
systematically approaching 2D from a sequence of quasi-1D 
ladder models\cite{2legDBL}.
The ladder discretization of the transverse momentum cuts through the 
2D surface, leading to a quasi-1D descendant state with a set of 
low-energy modes whose number grows with the number of legs and 
whose momenta are inherited from the 2D surfaces.   
These quasi-1D descendant states can be accessed in a controlled fashion
by analyzing the 1D ladder models using numerical and analytic 
approaches (ED, DMRG, VMC together with bosonization and gauge theory).
These multi-mode quasi-1D liquids constitute a new and previously 
unanticipated class of quantum states interesting in their own right.
But more importantly they carry a distinctive quasi-1D ``fingerprint" 
of the parent 2D quantum fluid state.

The power of this approach was demonstrated in a recent 
paper\cite{2legDBL} where we studied a new Boson-ring model on a 
two-leg ladder and mapped out the full phase diagram using DMRG and ED, 
supported by variational wavefunction and gauge theory analyses.
Remarkably, even for a ladder with only 2-legs, we found compelling 
evidence for the quasi-1D descendant of the 2D DBL phase.  
This new quasi-1D quantum state possessed all of the expected 
signatures reflecting the parent 2D Bose surface.

It will be most interesting to search for analogous 2D spin metal phases
in models possessing SU(2) spin symmetry.  Particularly promising are 
``weak Mott insulators" which are located in close proximity to the 
metal-insulator transition, such as the organic triangular lattice 
Mott insulator $\kappa$-(ET)$_2$Cu$_2$(CN)$_3$ which appears to exhibit 
a spin liquid ground state.
In these systems significant local charge fluctuations induce 
multi-spin ring exchange processes which tend to suppress magnetic or 
other types of ordering.  Several authors have proposed that a 
mean field state with a Fermi surface of spinons is an appropriate 
starting point\cite{Shimizu03, ringxch, SSLee}.
A preliminary analysis of the Heisenberg plus 4-site ring exchange
spin Hamiltonian on the $2-$leg triangular strip (using DMRG, ED, VMC, 
and gauge theory) is indicating strong evidence for the anticipated 
ladder descendant of the spinon Fermi sea state over a large swath 
of the phase diagram\cite{zigzag}.
It should be possible to extend this study to $3$ and $4-$leg 
triangular strips.  One could also study the half-filled Hubbard model 
on triangular strips to see if the quantum state just on the 
insulating side of the Mott transition is descended from this 2D 
spin-metal phase.  
A 2D non-Fermi liquid phase of itinerant electrons that has singular 
surfaces which violate Luttinger's theorem (surfaces with the ``wrong" 
volume, or perhaps even arcs), should also be accessible by 
systematically approaching 2D from a sequence of quasi-1D ladder models.
In this case the momenta of the low energy quasi-1D modes will likewise 
violate Luttinger's theorem (which is valid for a ``conventional" 
$N-$band Luttinger liquid).

\section{Acknowledgments}
This work was supported by
DOE grant DE-FG02-06ER46305 (DNS), 
the National Science Foundation through grants 
DMR-0605696 (DNS) and DMR-0529399 (MPAF), 
and the A.~P.~Sloan Foundation (OIM).



\end{document}